\documentclass[runningheads]{llncs}
\usepackage{graphicx}
\usepackage{amsmath,amssymb} 
\usepackage{color}
\usepackage{array}
\usepackage{subfig}
\usepackage{float}
\usepackage{multirow}
\usepackage{caption}

\begin{document}
\pagestyle{headings}
\mainmatter

\def\ACCV20SubNumber{771}  

\title{CS-MCNet:A Video Compressive Sensing Reconstruction Network with Interpretable Motion Compensation} 
\titlerunning{CS RECONSTRUCTION WITH INTERPRETABLE MC}
%
\author{Bowen Huang\inst{1} \and
Jinjia Zhou\inst{2,3}\and
Xiao Yan\inst{1} \and
Ming'e Jing\inst{1}\and
Rentao Wan\inst{1}\and
Yibo Fan\inst{1}}
\authorrunning{B.Huang, J.Zhou, et al.}
%
\institute{State Key Laboratory of ASIC and System, Fudan University, Shanghai, China\\
\email{fanyibo@fudan.edu.cn}\\
\url{}\and
Graduate School of Science and Engineering, Hosei University, Tokyo\and
JST, PRESTO, Japan\\
}

\maketitle

\begin{abstract}


In this paper, a deep neural network with interpretable motion compensation called CS-MCNet is proposed to realize high-quality and real-time decoding of video compressive sensing. Firstly, explicit multi-hypothesis motion compensation is applied in our network to extract correlation information of adjacent frames(as shown in Fig. \ref{fig1}), which improves the recover performance. And then, a residual module further narrows down the gap between reconstruction result and original signal. The overall architecture is interpretable by using algorithm unrolling, which brings the benefits of being able to transfer prior knowledge about the conventional algorithms. As a result, a PSNR of 22dB can be achieved at 64x compression ratio, which is about $4\%$ to $9\%$ better than state-of-the-art methods. In addition, due to the feed-forward architecture, the reconstruction can be processed by our network in real time and up to three orders of magnitude faster than traditional iterative methods.
\end{abstract}

\begin{figure*}
  \centering
  \includegraphics[width=\linewidth]{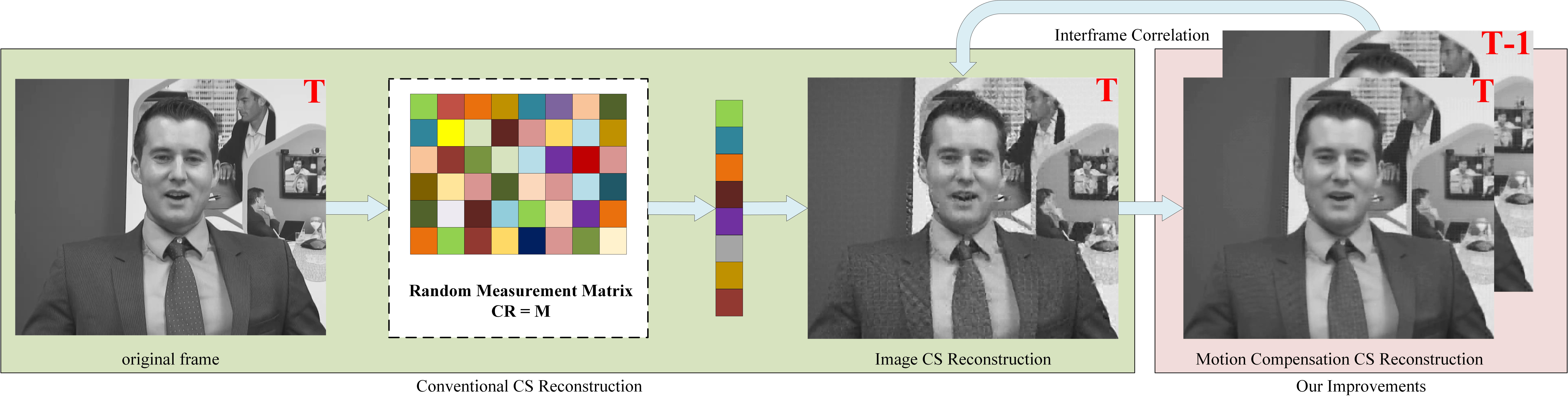}
  \caption{{\it Video CS reconstruction with explicit multi-hypothesis motion compensation}: CS measurement is acquired with random measurement matrix and preliminary result is gotten by using image CS reconstruction. Multi-hypothesis motion compensation is added to extract correlation information of adjacent frames, which improves recover performance. The reconstruction can be processed in real time.}
  \label{fig1}
\end{figure*}

\section{Introduction}

Traditional image or video compression methods, such as JPEG and H.265, compress the data after the measurement. However, compressive sensing, firstly introduced by Candes, Tao and Donoho\cite{4016283}\cite{1614066} in 2006, allows compression in the sensing process, {\it i.e.} sampling part of the signal instead of the entirety. It has been shown that if the target signal has transform sparse properties, {\it i.e.}being sparse in a transform domain, then it can be recovered from sample less than the Shannon-Nyquist sampling criterion requires\cite{4472247}. Suppose the target signal is $x \in C^{N}$, CS incorporates the compression into acquisition with a measurement matrix $\Phi \in C^{M\times N}$, where $M << N$:

\begin{equation}
y = \Phi \cdot x\;.
\label{eq1}
\end{equation}

Here y is the measurements. 

Even though traditional compression methods provides higher compression ratio and more mature system in some cases, the characteristic of simultaneous sensing and compression of compressive sensing requires very different encoder and decoder, which is of great importance to some specific areas, such as medical imaging systems, high frame rate video systems, and multimedia data compression.

The compression ratio CR can be defined as $CR = \frac{M}{N}$. CS reconstruction should be 'sparse', {\it i.e.} the original signal can be represented as $\hat{x} = \Psi \cdot x$, where $\Psi$ is called the sparsity basis and $\hat{x}$ the sparse representation of x. While natural images and video are difficult to achieve true sparsity, compressive sensing allows for approximate sparsity as well. Besides, CS reconstruction should also obey restricted isometry property, or RIP\cite{4472240}\cite{doi:10.1002/cpa.20124}. It has been proved that RIP rule is equal to measurement basis $\Phi$ and sparsity basis $\Psi$ being mutually incoherent\cite{Baraniuk2008ASP}. Thus, random matrix is commonly chosen in CS measurement. In addition, structured random matrix(SRM) can also meet the requirements and provide additional benefits, such as preserving information or reducing computation and memory consumption.(e.g., \cite{8551508}).

\subsection{Related Works}
In the recent decades, many methods have been proposed to solve the CS reconstruction problem\cite{6738005},\cite{doi:10.1002/cpa.20042},\cite{4907073},\cite{7457256}. For image reconstruction of compressive sensing, the algorithm of traditional transformation(e.g., wavelets domain\cite{4907073},\cite{QU2012964})runs fast but with low accuracy. Methods that rely on complex sparsity, such as dictionary learning methods\cite{7337391}generally have better reconstruction performance but lower computational speed. Furthermore, while most research efforts for image CS problems can be directly applied to video CS tasks, they fail to take advantage of the correlation between adjacent frames in a video sequence.

For video CS recovery, in\cite{6738005}, the authors use Gaussian mixture model(GMM) to recover high-frame-rate videos, and the reconstruction can be efficiently computed as an analytical solution. In\cite{5749476}, the authors propose a motion-compensation and block-based method MC-BCS-SPL, which estimate a motion vector from a reference frame and the under-reconstruction frame can then get prediction to improve recover performance. In general, these traditional methods focus on the design of different priors, transformations and sparsity constraints. However, these methods are usually difficult to determine the hyperparameters, such as thresholds or number of iterations, and due to their computational complexity, they can not perform real-time rebuild.

Driven by the powerful learning capabilities of neural networks, a number of DNN-based approaches have been applied\cite{8354291},\cite{YAO2019483},\cite{ILIADIS20189},\cite{7780424},\cite{8578294}. In\cite{8578294}, the authors cast ISTA into deep network form and develope an effective strategy to solve the proximal mapping associated with the sparsity-inducing regularizer using nonlinear transforms. In\cite{ILIADIS20189}, the authors propose a fully-connected neural network to reconstruct video temporal CS measurement, and a repetitive pattern measurement mask is proposed to make such a task practical. In\cite{8354291}, the authors propose a network named "CSVideoNet". The network combines a multi-rate CNN and a synthesizing RNN to improve the trade-off between compression ratio and spatial-temporal resolution of the reconstructed videos. Compared with iterative algorithms, these feed-forward methods significantly reduce time consumption. However, the structure of these networks is often empirically determined and is ambiguous as a black box, which brings difficulty in making targeted improvements. 

To resolve the conflict between interpretability and speed of reconstruction, a technique called algorithm unrolling has been applied recently. The technique was proposed by Gregor et al.\cite{10.5555/3104322.3104374}, and builds neural network by unfolding an iterative optimization algorithm to be a hierarchical architecture, which provides a principled framework by expressing traditional iterative algorithms as neural networks, and offers promise in developing interpretable network. There are several networks using algorithm unrolling to solve CS reconstruction\cite{8578294},\cite{8550778}, but they are developed for image CS tasks rather than specifically for video CS tasks.

In this work, we develop a network, called CS-MCNet, that attempts to use inter-frame information to improve the reconstruction quality of video CS measurements. By mapping the iterative algorithm MC-BCS-SPL into non-iterative neural network, all of CS-MCNet's block is designed to correspond to an iteration in MC-BCS-SPL. By end-to-end training, our network can learn all parameters efficiently.

\subsection{Contribution}
Our main contributions are summarized as follows.1)We use neural network modules to replace the optimization steps in traditional model-based approaches and implement them in a simple form that is easy to realize quickly. 2)We propose a multi-hypothesis motion compensation structure. The module exploits the similarity between neighboring frames to improve the reconstruction quality. To the best of our knowledge, it is the first work that explicitly uses motion compensation for video CS reconstruction in deep neural network. 3)We employ a residual module in the network to further improve performance, and this structure also facilitates the training of deeper neural networks. With these improvements, our work outperforms previous work in terms of both reconstruction quality and computational cost.

\section{Methodology}
\begin{figure*}
\centering
\includegraphics[width=\linewidth]{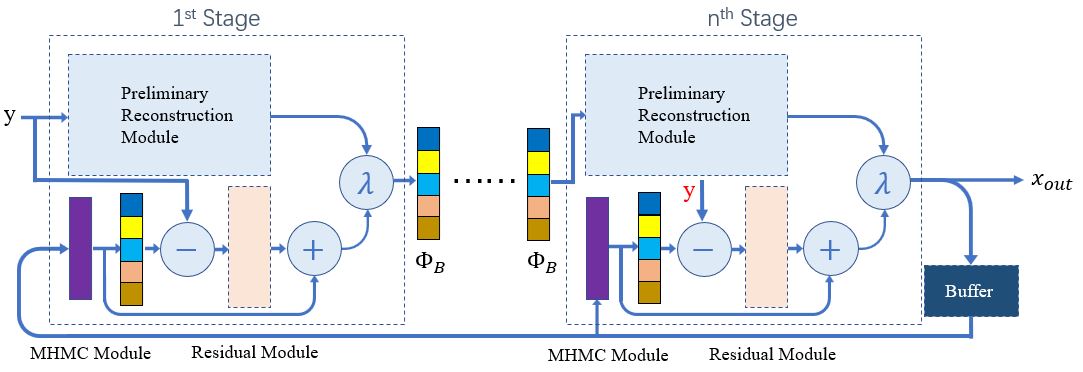}
\caption{The overall architecture of the proposed network. The input $y$ is acquired from video frames by compressive sensing. The network is composed of several stages and the reconstruction is performed by three modules in each stage, which is corresponding to the blue, purple and pink modules in the figure. A buffer is designed to store recovery results of one frame and offer reference for the reconstruction of the next frame.}
\label{fig2}
\end{figure*}

By taking advantage of the merits of model based and DNN-based CS methods, CS-MCNet maps the optimization steps into a deep network architecture consisting of a fixed number of stages, each of which is designed to correspond to one iteration in the MC-BCS-SPL algorithm. The overall architecture of CS-MCNet is shown in Fig. \ref{fig2}. The proposed CS-MCNet consists of an encoder (sensing matrix) and a decoder. The encoder performs simultaneous sampling and compression. The decoder consists of several stages, each of which is divided into three parts. Firstly, the decoder roughly recovers input measurements to get a preliminary result. Secondly, it get prediction from a single reference frame. Then the prediction is measured and subtracted from original measurements to get the residual. Thirdly, the residual measurements are recovered and the result is added to the prediction. The output is derived from combining the preliminary result and residual reconstruction linearly.  We will introduce each module in the following subsections.

\subsection{Priliminary reconstruction module}
The first hidden layer is a fully-connected layer that would provide 3D signal from 2D compressed measurements. Several papers have shown that CNNs can achieve superior performance on CS reconstruction problems compared with simple optimization-based algorithms.\cite{7780424},\cite{ILIADIS20189},\cite{YAO2019483}, thus we use CNN to get the preliminary reconstruction result. Typical CNN architectures used to do recognition, classification, and segmentation are not suitable to the reconstruction problem here. The goal of CNNs in our network is to retain as much detail as possible and need to recover pixels that do not exist based on known information. Therefore, we eliminated the pooling layer, which causes information loss. 

\begin{figure*}
\centering
\includegraphics[width=\linewidth]{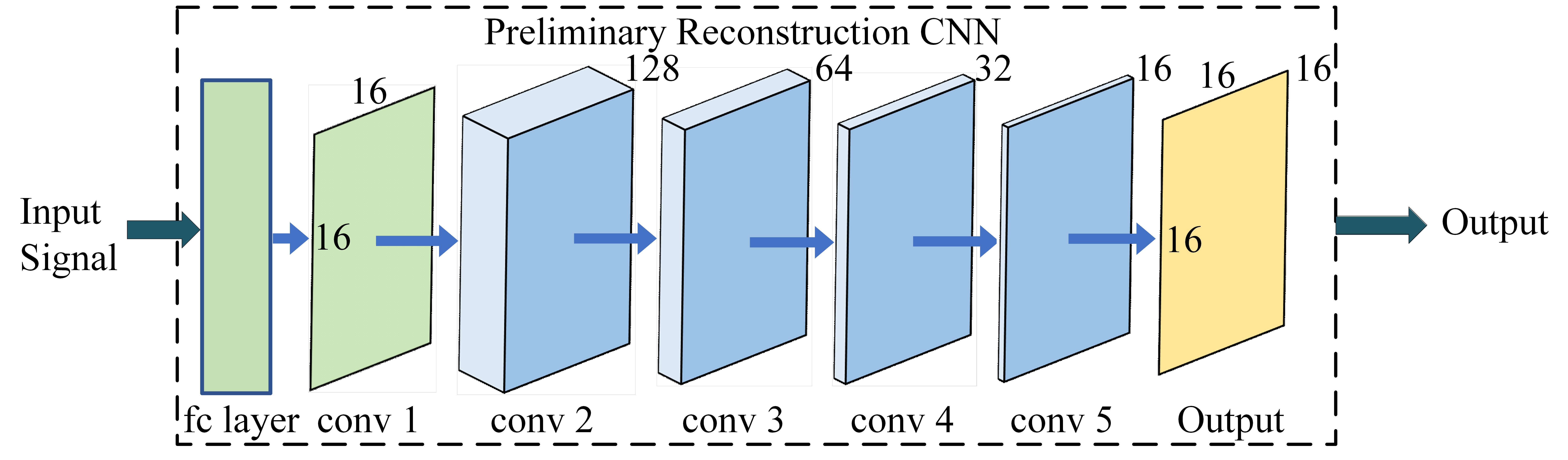}
\caption{Detailed architecture and corresponding parameter of the preliminary reconstruction CNN.}
\label{fig3}
\end{figure*}

To reduce the size of parameters and simplify the network architecture, we use video blocks as input and set the block size to $16\times16$. The convolutional layers, each of which is followed by a ReLU layer except for the final layer, are carefully designed to get amenable recovery performance. All feature maps are the same size as the reconstructed video block, and the number of feature maps is monotonically reduced. The detailed structure of the CNN is shown in Fig. \ref{fig3}. This process resembles the sparse coding stage in CS, where a subset of dictionary atoms are combined to form an estimation of the original input.  To improve final reconstruction performance, we pre-train the CNN before training the whole CS-MCNet, since the path is long from the input to the output of the whole net and pre-training can help prevent the vanishing gradient problem\cite{erhan2010why}.

\subsection{Multi-hypothesis motion compensation module}
Traditional video compression algorithms have long exploited motion compensation to improve video-coding quality\cite{10.5555/2843012},\cite{1094950}; In consideration of bit rate at the encoder side, these techniques use single hypothesis in order to limit the amount of motion vector. However, this limitation doesn't exist since the motion compensation is all calculated at the decoder side of the system. Thus, multi-hypothesis motion compensation can be considered to combine all available best assumptions in the reference frame. A MH CS reconstruction can be represented as an optimal linear combination of all possible reference blocks:

\begin{equation}
w_{t,i} = arg \min\limits_{w}||x_{t,i} - H_{t,i}w||_{2}^{2}\;.
\label{eq2}
\end{equation}
\begin{equation}
\tilde{x}_{t,i} = H_{t,i}w_{t,i}\;.
\label{eq3}
\end{equation}

Here, the subscript "t" and "i" represents the index of frame in the video and the index of block in the frame, and $H_{t,i}$ is a matrix of dimensionality $B^{2}\times K$ consisted of rasterizations of the possible blocks within the search window in the reference frame, and $K = |H_{t,i}|$. In this context, $w_{t,i}$ represents the linear combination of the columns of $H_{t,i}$; The solution of this optimization can be calculated as a least-squares(LSQ) problem\cite{5749477}.

The proposed network in this paper uses MH motion compensation to improve the recovery performance. Unlike traditional optimization-based solutions to the LSQ problem, we use fully-connected layers to learn the optimal parameters. Due to the similarity of adjacent frames, this MH motion compensation module can be trained appropriately to produce accurate predictions of motion and the recovery quality can be improved by the aggregation of motion and spatial visual features. 

To get the reference frame, we design a buffer to store the reconstructed video blocks, as shown in Fig. \ref{fig2}. For the sake of simplicity, we choose to do reconstruction after the reference frame is completely reconstructed. However, the search window actually only involves part of the reference frame, and therefore we can reduce the size of the buffer by carefully designing the rebuilding order. In\cite{8354291}, the authors use LSTM network to do temporal reference, which is similar to motion compensation. However with the experiment, we prove that the utilization of explicit motion compensation module outperforms the RNN based methods and decrease the model size simutaneously. 

\subsection{Residual reconstruction module}
With the prediction of MH motion compensation, we introduce the residual reconstruction module to further narrow down the gap between $x_{reconstruction}$ and $x$. The output of residual learning is fused with the output of preliminary reconstruction module as the final result.

\begin{figure}
\centering
\includegraphics[width=0.8\linewidth]{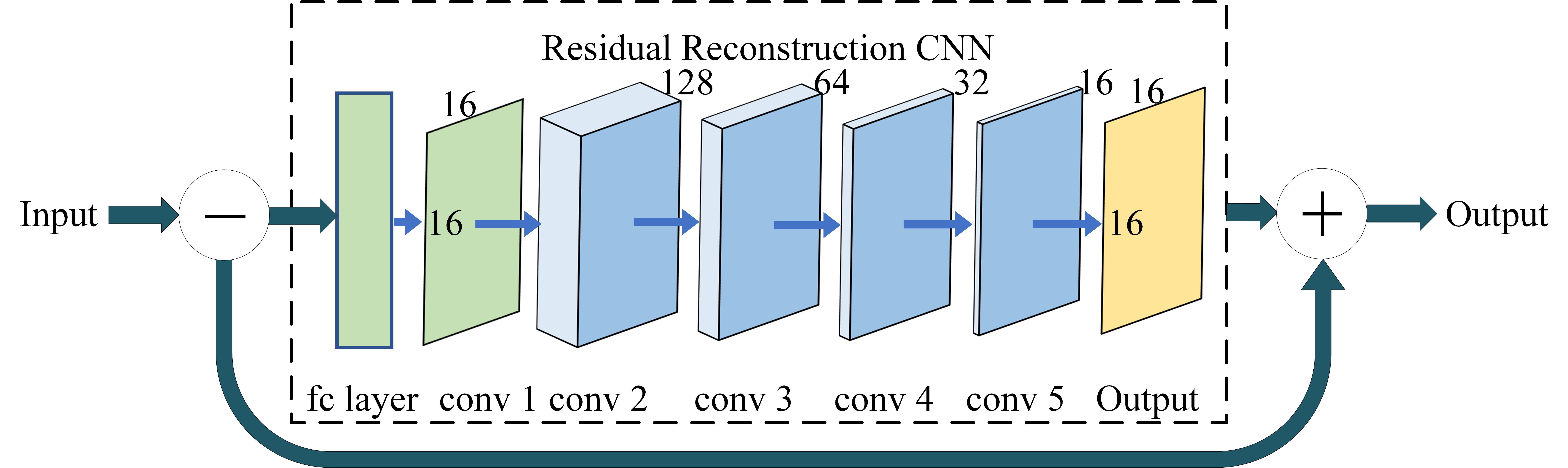}
\caption{Detailed architecture and corresponding parameter of the residual reconstruction module.}
\label{fig4}
\end{figure}

We get the residual signal $d$ by measuring the result of MH motion compensation and subtracting it from the original measurements $y$,
\begin{equation}
d_{i} = y_{i} - \Phi_{B}x_{mc,i}\;.
\label{eq4}
\end{equation}

According to\cite{7780459}, the convolutional layers in residual module could be easily trained to approximate the residual, most of which is zero. As shown in Fig. \ref{fig4}, the residual reconstruction module contains one fully-connected layer and five convolution layers. The fully-connected layer has the same function of recover 3D volume signal from 2D measurements as in preliminary reconstruction module. The rest has decreasing number of feature maps and holds the size of $16x16$. All convolutional layers is followed by a ReLU layer except the last one.

\subsection{Learning Algorithm}

Given the training data pairs $\{(x,x_{ref}\}_{i}$, CS-MCNet firstly gets measurements $y_{i}$ with sensing matrix. Then $y_{i}$ serves as input of the decoder and generates the reconstruction result. We wanted to reduce the discrepancy between the raw measurements and the MH motion compensation measurements, {\it i.e.}the residual signal. Therefore, the loss function for CS-MCNet is designed as follows:

\begin{equation}
L_{total}(\omega,b) = L_{err} + \lambda L_{mc}\;.
\label{eq5}
\end{equation}
\begin{equation}
\left\{  
\begin{array}{lr} 
L_{err} = \frac{1}{2N}\Sigma_{i}^{T}||f(y_{i};\omega,b) - x_{i}||_{2}^{2}\;, &\\
L_{mc} = \frac{1}{2N}\Sigma_{i}^{T}||y_{i} - \Phi_{B}x_{mc,i}||_{2}^{2} \;.      
\end{array}  
\right.
\label{eq6}
\end{equation}

where $\lambda$ is the scale factor to control the influence of motion compensation on the total loss. Determined by experiment, we set $\lambda$ to be 0.5 during training to get the best performance. 

We choose MSE to calculate the loss, which is a commonly used metric to quantitatively evaluate recovery quality. The proposed framework can also be adapted to other loss functions. Adam optimizer with default parameters is chosen to optimize the proposed network. 
 
\section{Experiment}
We compare our methods with state-of-the-art approaches, including iterative optimization based methods and DNN based methods. For fairness, we set the block size of $16\times16$ and retrained the reference networks with our self-built dataset. It should be emphasized that we rewrote CSVideoNet with Pytorch, whose original code was implemented by Torch. We took parameters from the original model files provided by the authors, however we cannot guarantee that it will achieve the same performance as the original code. Furthermore, to prove the advantage of exploiting motion compensation, image specific methods are also included. Both noiseless and noise measurements are tested and we also discuss the performance of our proposed network under different network parameters({\it i.e.} number of stages). Two metrics, peak signal-to-noise ration(PSNR) and structural similarity(SSIM) are used to evaluate the performance, and visualization of the results are provided.

\subsection{Implementation Details}
As there is no standard dataset designed for video CS, we use UCF-101 dataset\cite{Soomro2012UCF101AD} to build our own training dataset. UCF-101 dataset includes 13k clips and about 27 hours of video, which is collected from YouTube and is divided into 101 action classes. We extract only the luminance component of the extracted frames and crop the central $160x160$ patch from each frame. All of the patches are segmented into $16x16$ non-overlapping blocks. We randomly choose video sequences from UCF-101 dataset and finally get around 300,000 pairs of data for training and validation in total. 

Our model is implemented with PyTorch and all the experiments are performed on a workstation with an Intel Xeon CPU and a Nvidia GeForce RTX2080 GPU. Our networks are trained for 200 epochs with batch size of 400. We normalize the input pre-feature to zero mean and standard deviation one. We set the starting learning rate to 0.01 and divide the learning rate by 10 if the loss of the current epoch is greater than that of the previous epoch. Except for the last subsection, we use 4 cascaded stages in the following experiments.

\subsection{Comparison with the State-of-the-art}
\begin{figure}
\centering
\includegraphics[width=0.8\linewidth]{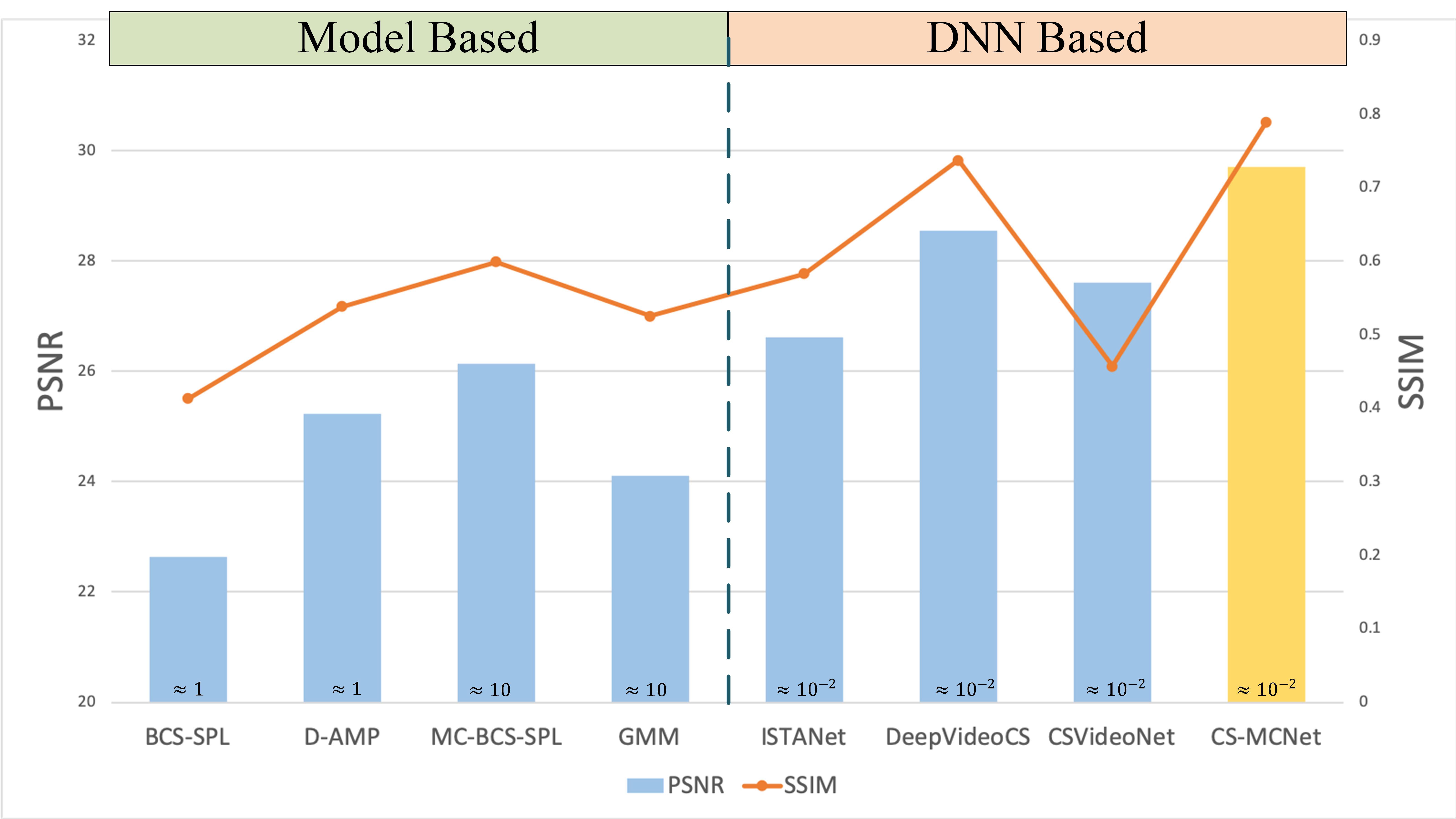}
\caption{Performance Comparison with different reference methods on test dataset. The time (at the bottom of the histogram) refers to the magnitude of the average time for reconstructing each frame.}
\label{fig5}
\end{figure}

We compare the reconstruction performance of our proposed method with several reference work of CS reconstruction\cite{5749476},\cite{7457256},\cite{6738005},\cite{8354291},\cite{8578294},\cite{ILIADIS20189}. The summarized information about all baseline approaches is listed in Table \ref{tab1}. All the methods reconstruct video blocks from its CS measurements independently, and the result of average PSNR, SSIM and time consumption for each method on the test dataset is reported in Fig. \ref{fig5}. To save training time, all methods are tested under CR of 16. From the results we can observe that CS-MCNet outperforms the reference method in terms of metrics and time consumption. Compared with conventional image CS algorithm BCS-SPL, D-AMP, and video CS algorithm MC-BCS-SPL, GMM, our DNN-based CS-MCNet benefits from learnable parameters and feed-forward architecture, and thus gets better reconstruction quality and uses less time. The similar DNN based methods DeepVideoCS, CSVideoCS and ISTANet either uses barely CNN or combines CNN and RNN to extract inexplicit motion features. In contrast to them, our work exploits explicit MH motion compensation, further improving the quality of the reconstruction and compressing the size of the model, which makes it easier and faster to train and deploy.

\setlength{\tabcolsep}{4pt}
\begin{table}[htbp]
\begin{center}
\caption{Classification and summary information for all reference methods and the proposed approach}
\label{tab1}
\begin{tabular}{>{\raggedright\arraybackslash}p{40 pt} |>{\raggedright\arraybackslash}p{60 pt} | >{\raggedright\arraybackslash}p{70 pt} | >{\raggedright\arraybackslash}p{120 pt}}
\noalign{\smallskip}
\hline
\noalign{\smallskip}
\multirow{3}{*}{Image CS} & \multirow{2}{*}{Model Based} & BCS-SPL\cite{5749476}     & block based CS with smooth projected Landweber \\
                          &                              & D-AMP\cite{7457256}          & Denoising-based approximate message passing\\
                          \noalign{\smallskip}
                          \cline{2-4}
                          \noalign{\smallskip}
                          & DNN Based                    & ISTANet\cite{8578294}     & CNN inspired by ISTA algorithm \\
\noalign{\smallskip}
\hline
\noalign{\smallskip}
\multirow{5}{*}{Video CS} & \multirow{2}{*}{Model Based} & MC-BCS-SPL\cite{5749476}  &  motion compensation block based CS \\
                          &                              & GMM\cite{6738005}         & Gaussian mixture model \\
                          \noalign{\smallskip}
                          \cline{2-4}
                          \noalign{\smallskip}
                          & \multirow{3}{*}{DNN Based}   & DeepVideoCS\cite{ILIADIS20189} & deep neural network with fully-connected layers \\
                          &                              & CSVideoNet\cite{8354291}  &a multi-rate CNN and a synthesizing RNN  \\
                          &                              & {\it CS-MCNet}    & proposed approach \\
\hline
\end{tabular}
\end{center}
\end{table}
\setlength{\tabcolsep}{1.4pt}

To further validate the advantages of using MH motion compensation, we do comparison with MC-BCS-SPL and ISTA under different CRs of 4,16 and 64. As shown in Table \ref{tab2}, CS-MCNet achieves relatively better performance, especially under high compression ratio. CS-MCNet outperforms MC-BCS-SPL both on reconstruction quality and time consumption, since its learnable parameters can be better optimized through end-to-end training, and GPU acceleration makes it three orders of magnitude faster than iterative methods. As for ISTANet, which also uses algorithm unrolling and has a similar residual block structure, the utilization of inter-frame information helps our network achieve better reconstruction quality, though the storage of reference frame requires extra storage space and time consumption.

\setlength{\tabcolsep}{4pt}
\begin{table}
\begin{center}
\caption{Performance comparison with MC-BCS-SPL\cite{5749476} and ISTANet\cite{8578294} under different CRs on test dataset.}
\label{tab2}
\begin{tabular}{>{\raggedright\arraybackslash}p{20 pt} | >{\raggedright\arraybackslash}p{30 pt} | >{\raggedright\arraybackslash}p{60 pt} >{\raggedright\arraybackslash}p{60 pt} >{\raggedright\arraybackslash}p{60 pt}}
\noalign{\smallskip}
\hline
\noalign{\smallskip}
CR & Metric & ISTANet\cite{8578294} & MC-BCS-SPL\cite{5749476} & {\it CS-MCNet:proposed}\\
\noalign{\smallskip}
\hline
\noalign{\smallskip}
4 & PSNR &\it{33.851} &31.067&33.35 \\
& SSIM &\it{0.953} &0.834 &0.918 \\
\noalign{\smallskip}
\hline
\noalign{\smallskip}
16 & PSNR &26.618 &26.141&\it{29.707}\\
& SSIM &0.583 &0.436&\it{0.789}\\
\noalign{\smallskip}
\hline
\noalign{\smallskip}
64 & PSNR &19.93 &19.79 &\it{21.45} \\
& SSIM &0.382 &0.288 &\it{0.528} \\
\hline
\end{tabular}
\end{center}
\end{table}
\setlength{\tabcolsep}{1.4pt}

\begin{figure*}
\centering
\begin{minipage}[b]{0.9\linewidth}
  \subfloat{
    \begin{minipage}[b]{0.3\linewidth}
      \centering
      \setlength{\abovecaptionskip}{0pt}
      \setlength{\belowcaptionskip}{0pt}
      \includegraphics[width=\linewidth]{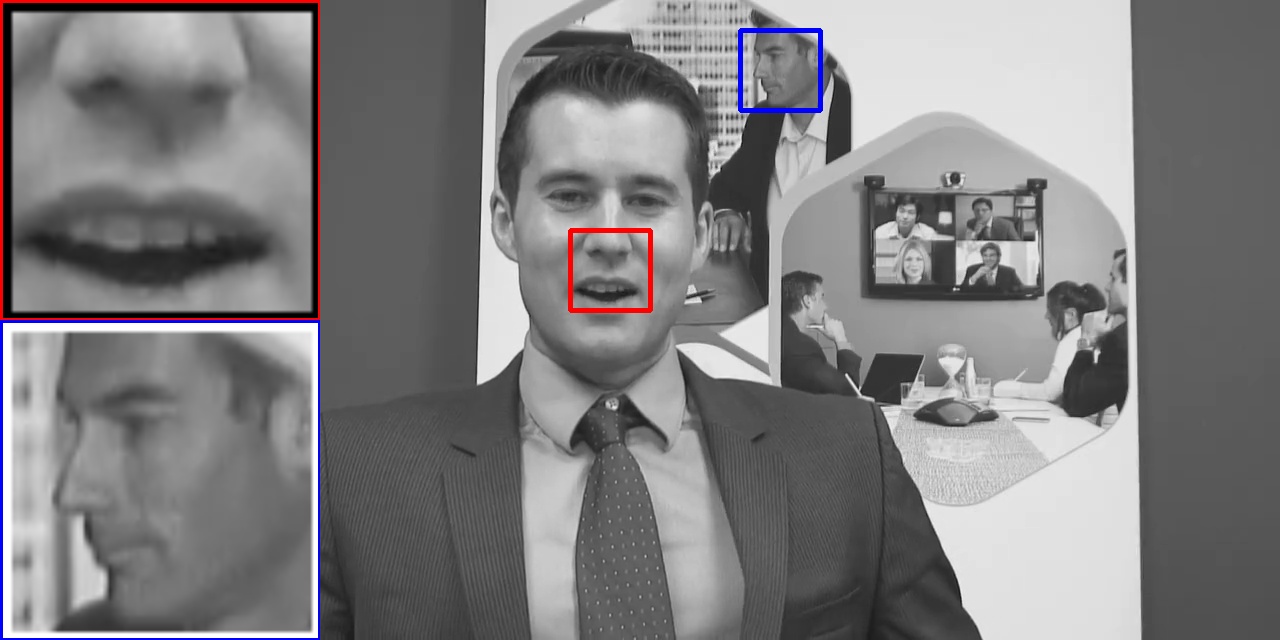}
      \caption*{Gound Truth\protect\\-/-}
      \includegraphics[width=\linewidth]{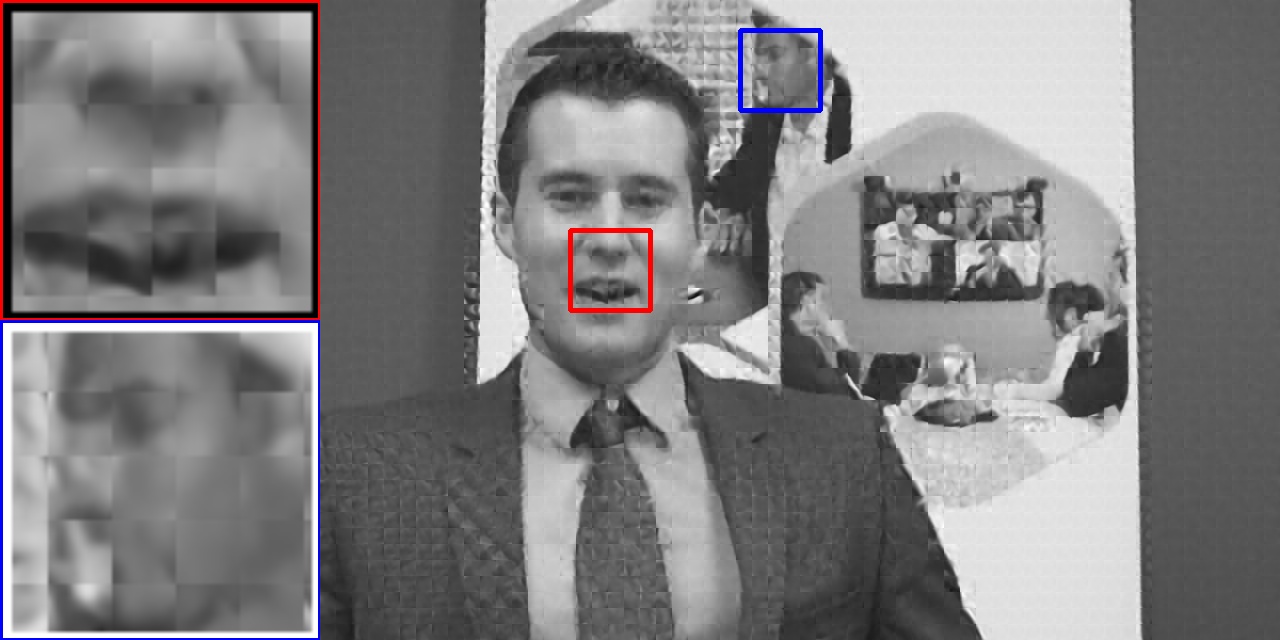}
      \caption*{ISTANet\cite{8578294}\protect\\30.375/0.871}
      \includegraphics[width=\linewidth]{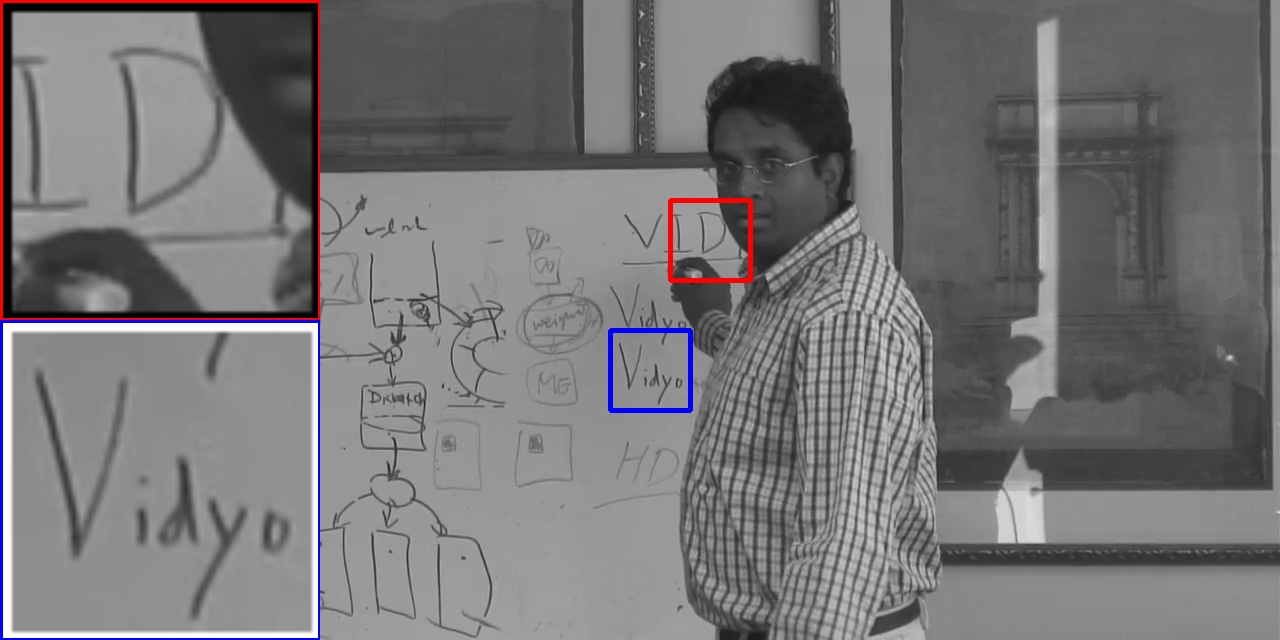}
      \caption*{Gound Truth\protect\\-/-}
       \includegraphics[width=\linewidth]{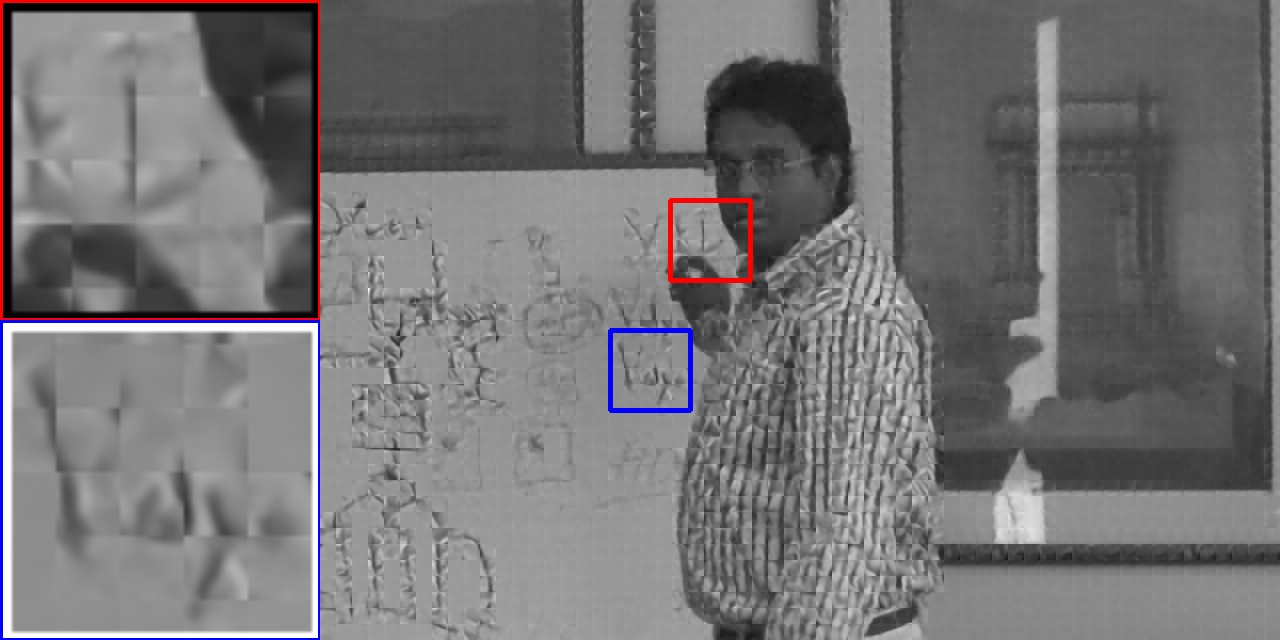}
      \caption*{ISTANet\cite{8578294}\protect\\26.621/0.806}
      \includegraphics[width=\linewidth]{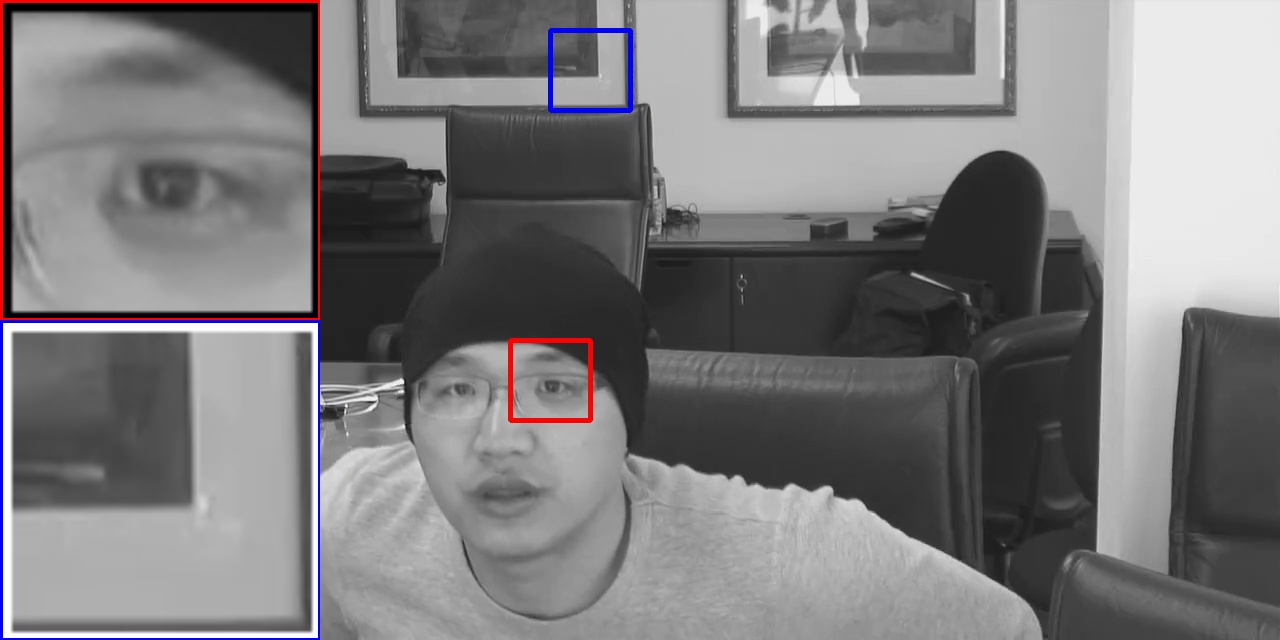}
      \caption*{Goudn Truth\protect\\-/-}
       \includegraphics[width=\linewidth]{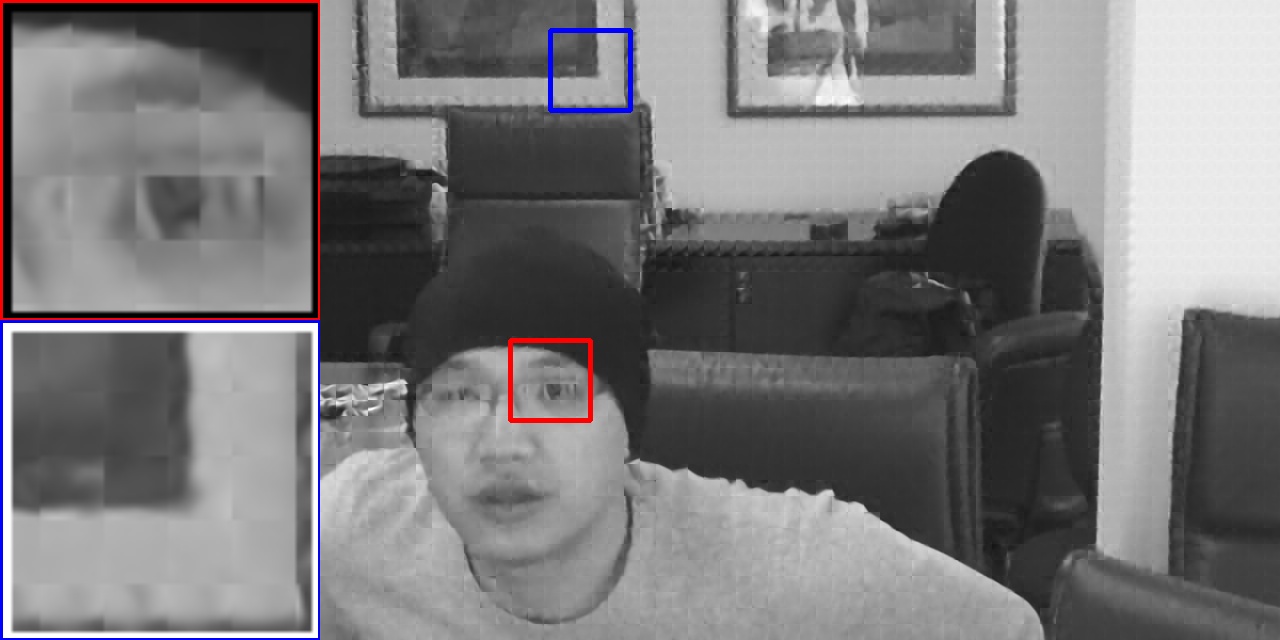}
      \caption*{ISTANet\cite{8578294}\protect\\28.592/0.865}
    \end{minipage}
  }
  \hfill
     \subfloat{
    \begin{minipage}[b]{0.3\linewidth}
      \centering
      \setlength{\abovecaptionskip}{0pt}
      \setlength{\belowcaptionskip}{0pt}
       \includegraphics[width=\linewidth]{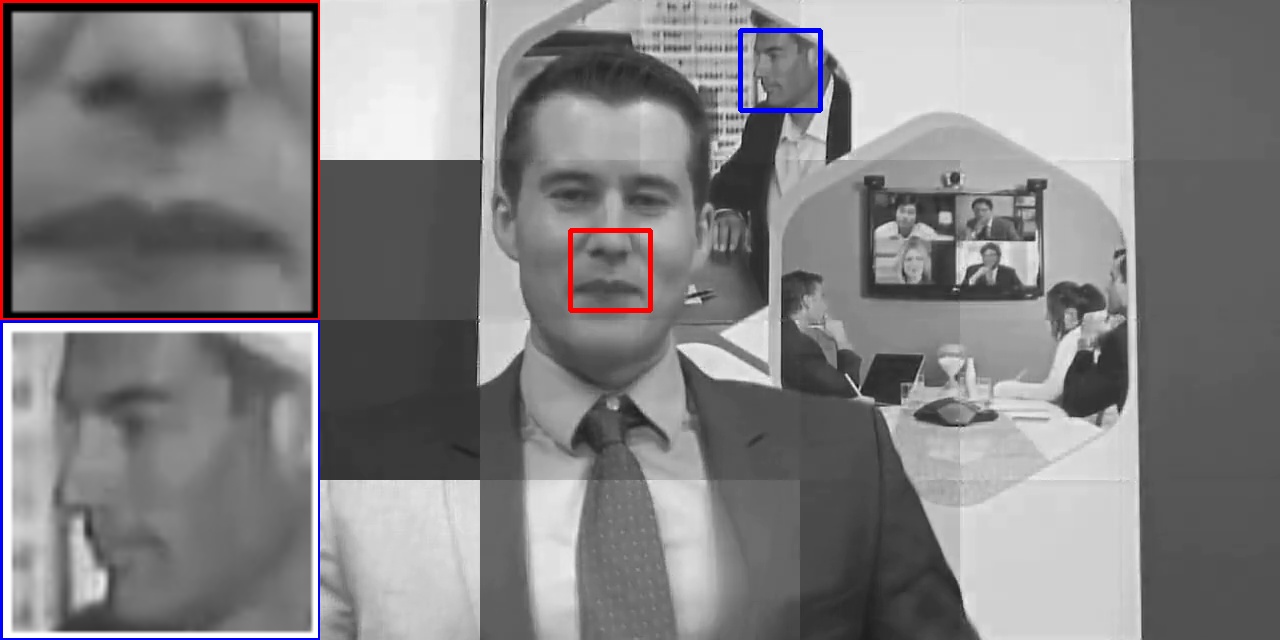}
      \caption*{DeepVideoCS\cite{ILIADIS20189}\protect\\10.104/0.791}
      \includegraphics[width=\linewidth]{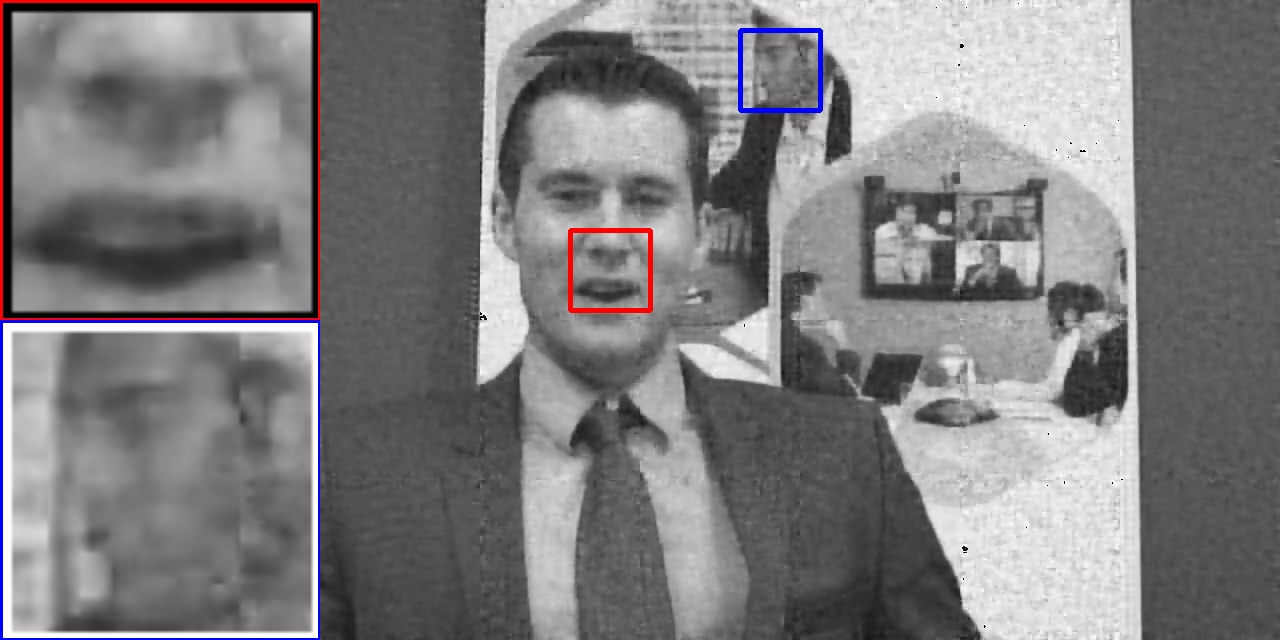}
      \caption*{CSVideoNet\cite{8354291}\protect\\29.603/0.812}
      \includegraphics[width=\linewidth]{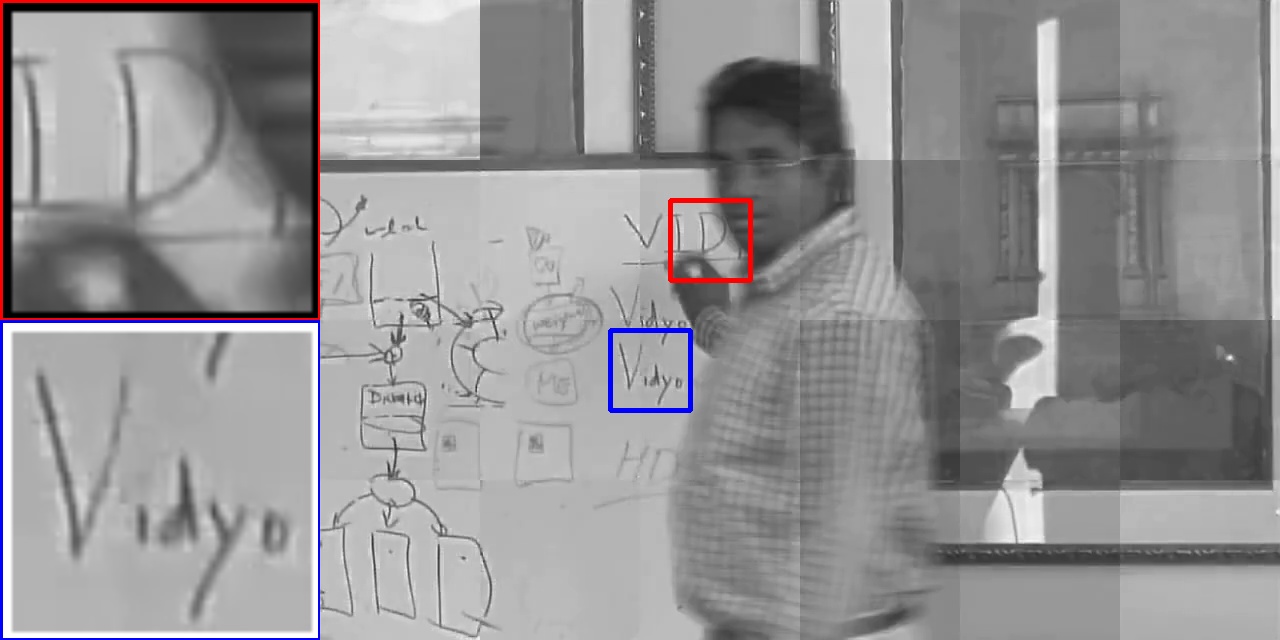}
      \caption*{DeepVideoCS\cite{ILIADIS20189}\protect\\14.608/0.797}
       \includegraphics[width=\linewidth]{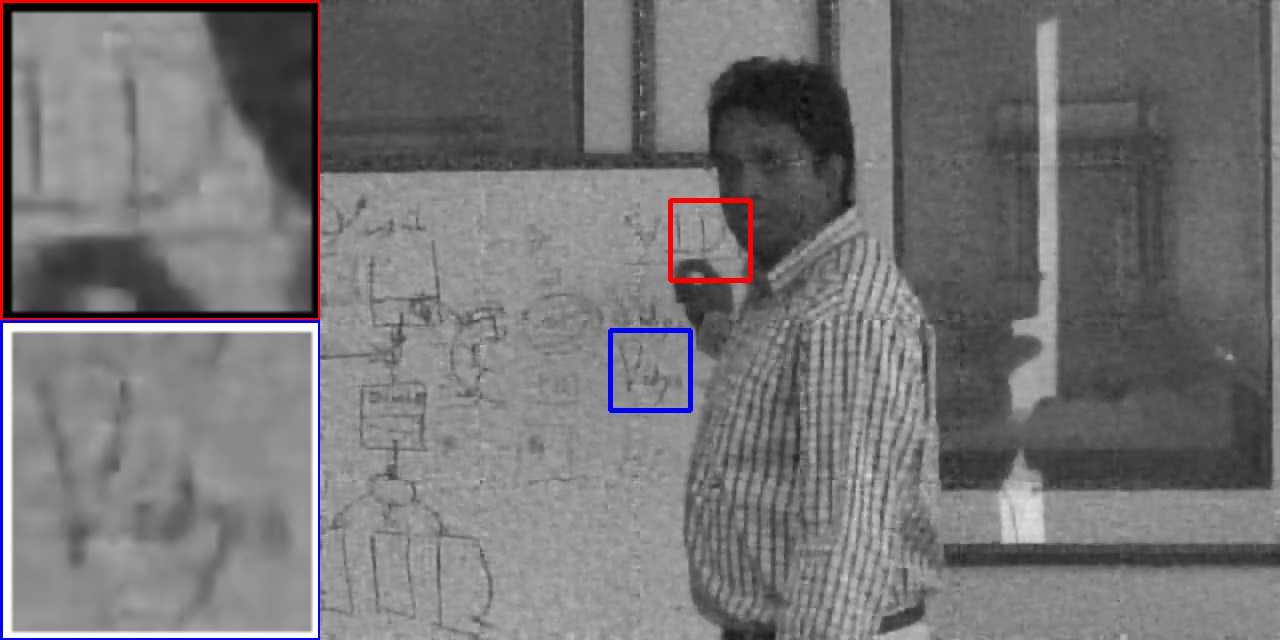}
      \caption*{CSVideoNet\cite{8354291}\protect\\25.713/0.728}
      \includegraphics[width=\linewidth]{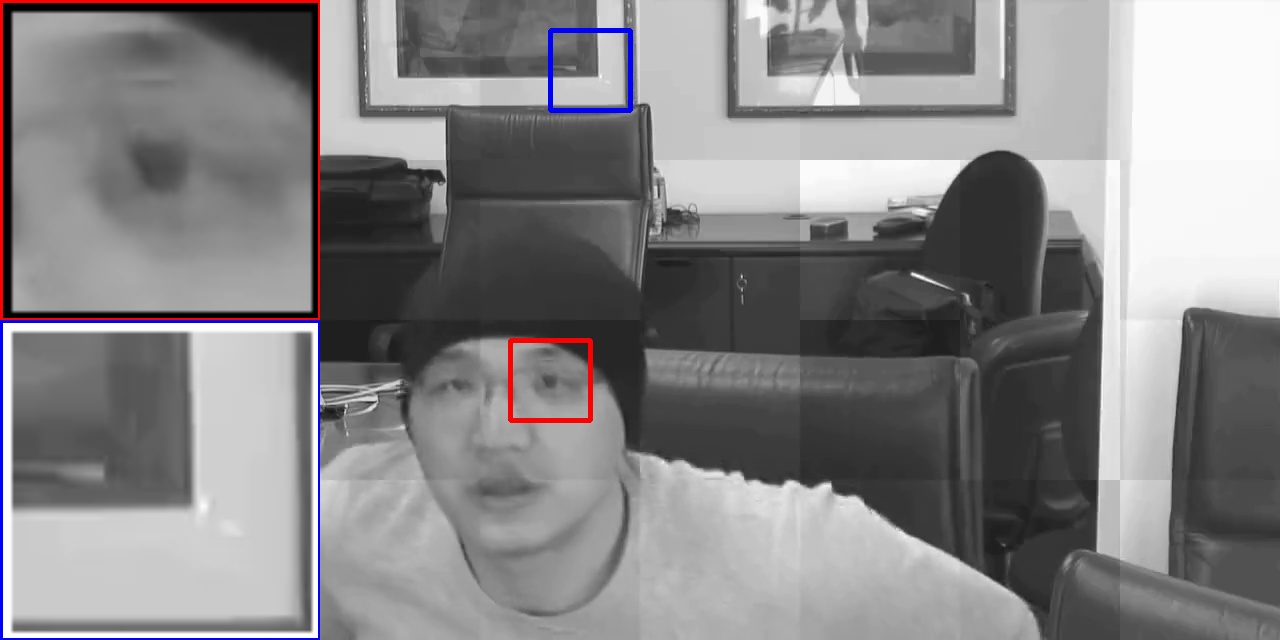}
      \caption*{DeepVideoCS\cite{ILIADIS20189}\protect\\20.501/0.917}
       \includegraphics[width=\linewidth]{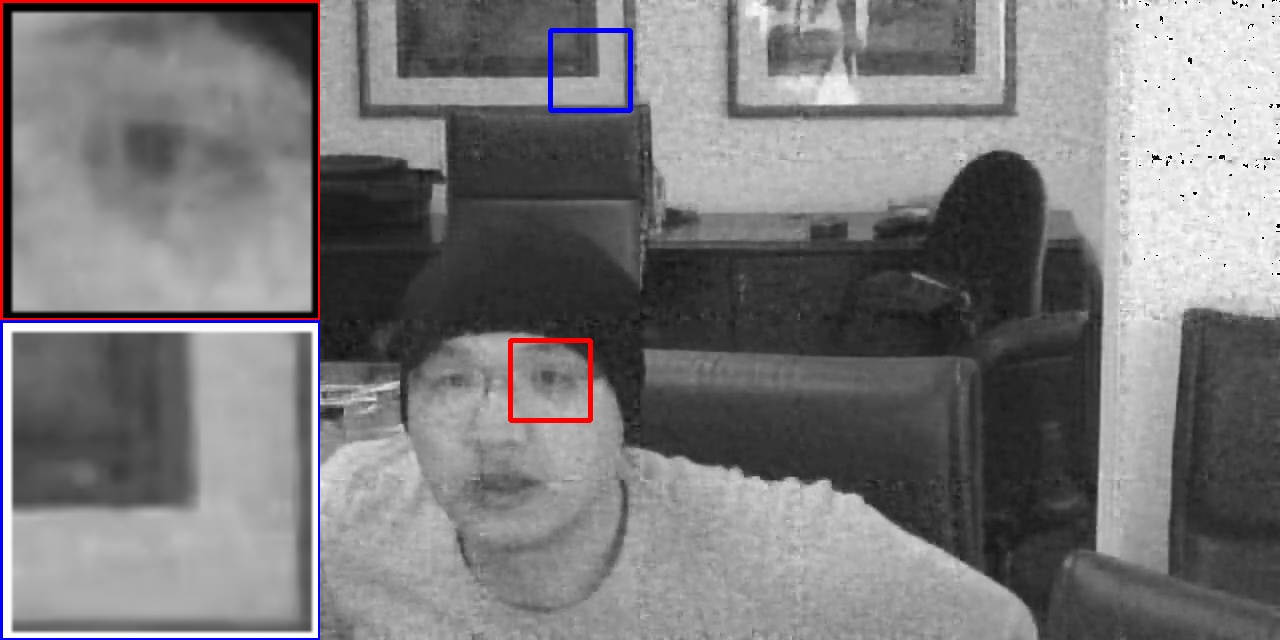}
      \caption*{CSVideoNet\cite{8354291}\protect\\25.790/0.772}
    \end{minipage}
  }
  \hfill
     \subfloat{
    \begin{minipage}[b]{0.3\linewidth}
      \centering
      \setlength{\abovecaptionskip}{0pt}
      \setlength{\belowcaptionskip}{0pt}
       \includegraphics[width=\linewidth]{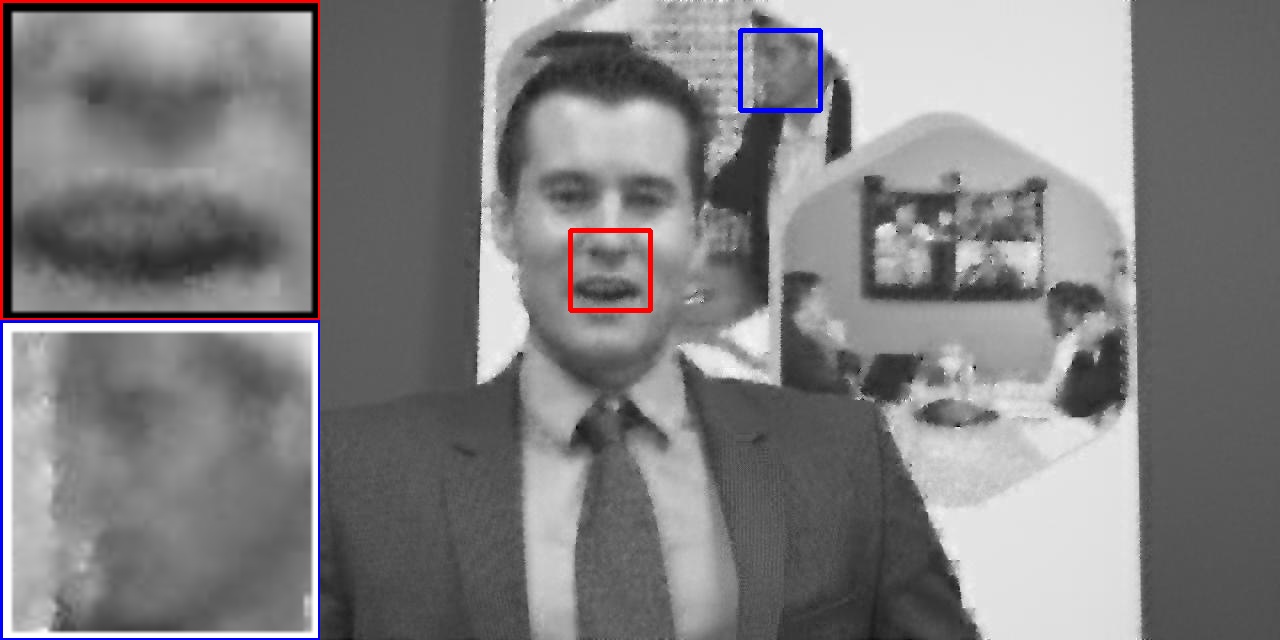}
      \caption*{MC-BCS-SPL\cite{5749476}\protect\\29.206/0.880}
      \includegraphics[width=\linewidth]{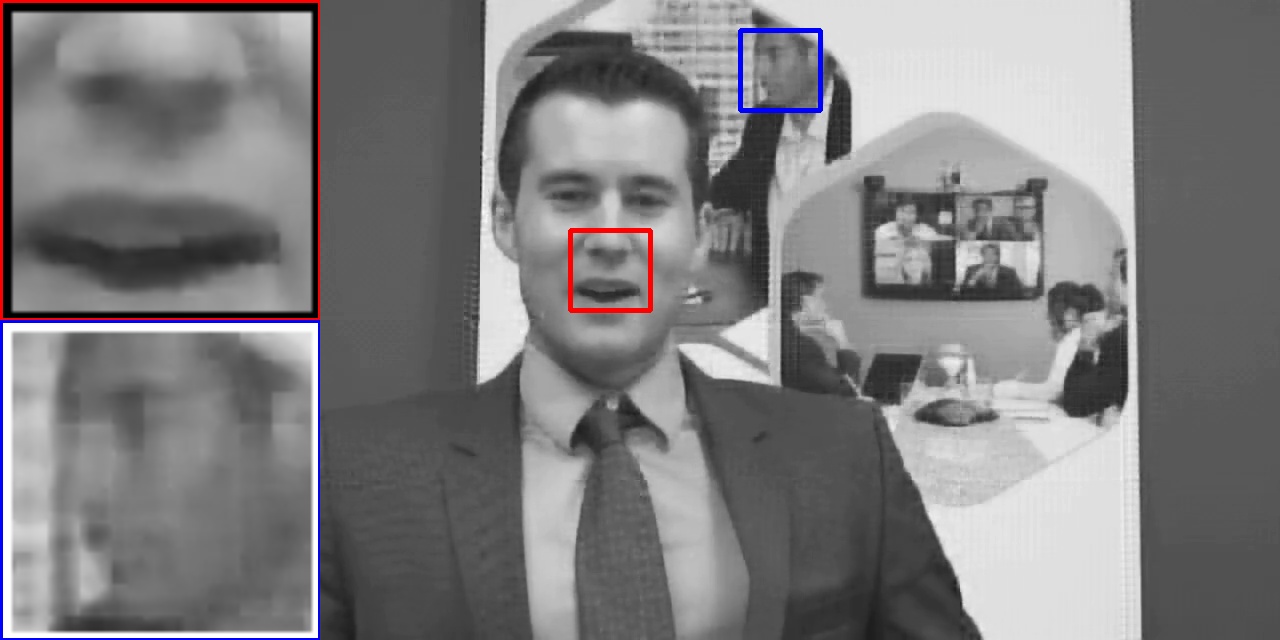}
      \caption*{CS-MCNet{\it (proposed)}\protect\\32.099/0.908}
      \includegraphics[width=\linewidth]{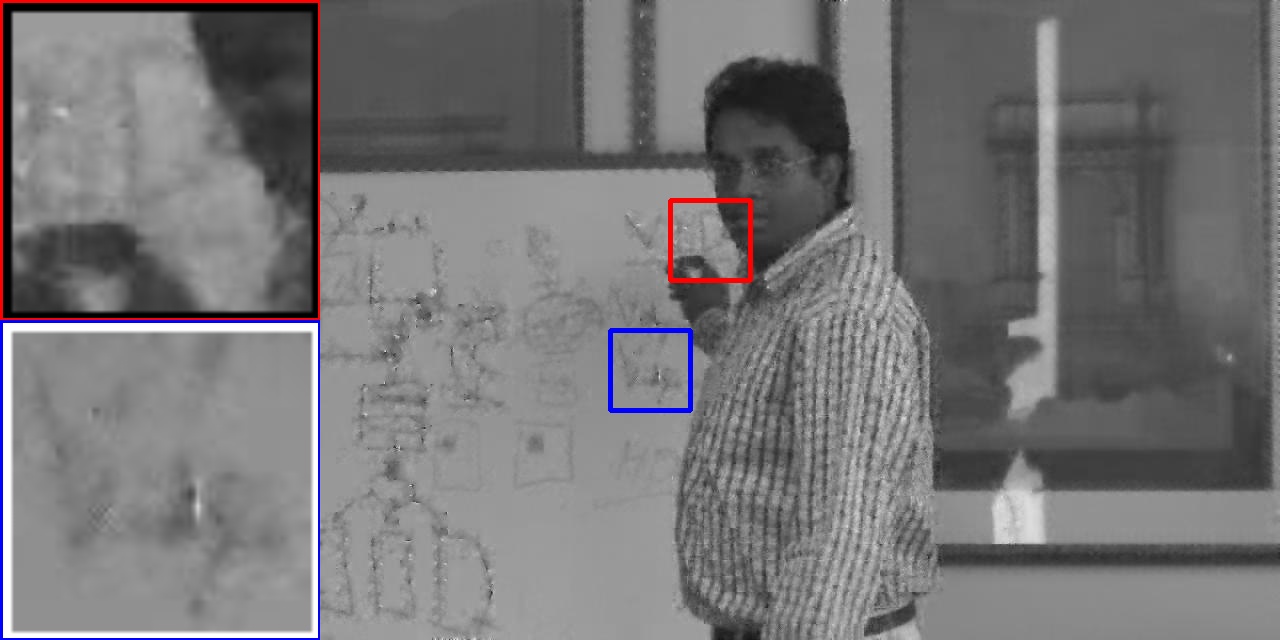}
      \caption*{MC-BCS-SPL\cite{5749476}\protect\\25.371/0.780}
       \includegraphics[width=\linewidth]{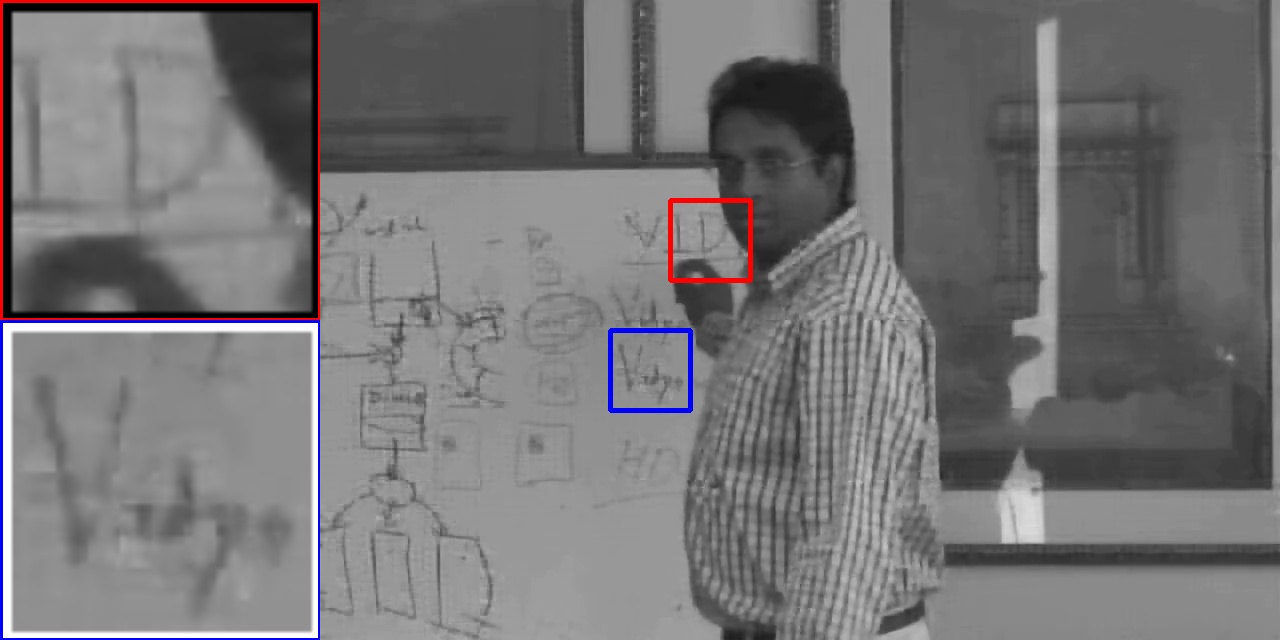}
      \caption*{CS-MCNet{\it (proposed)}\protect\\26.294/0.817}
      \includegraphics[width=\linewidth]{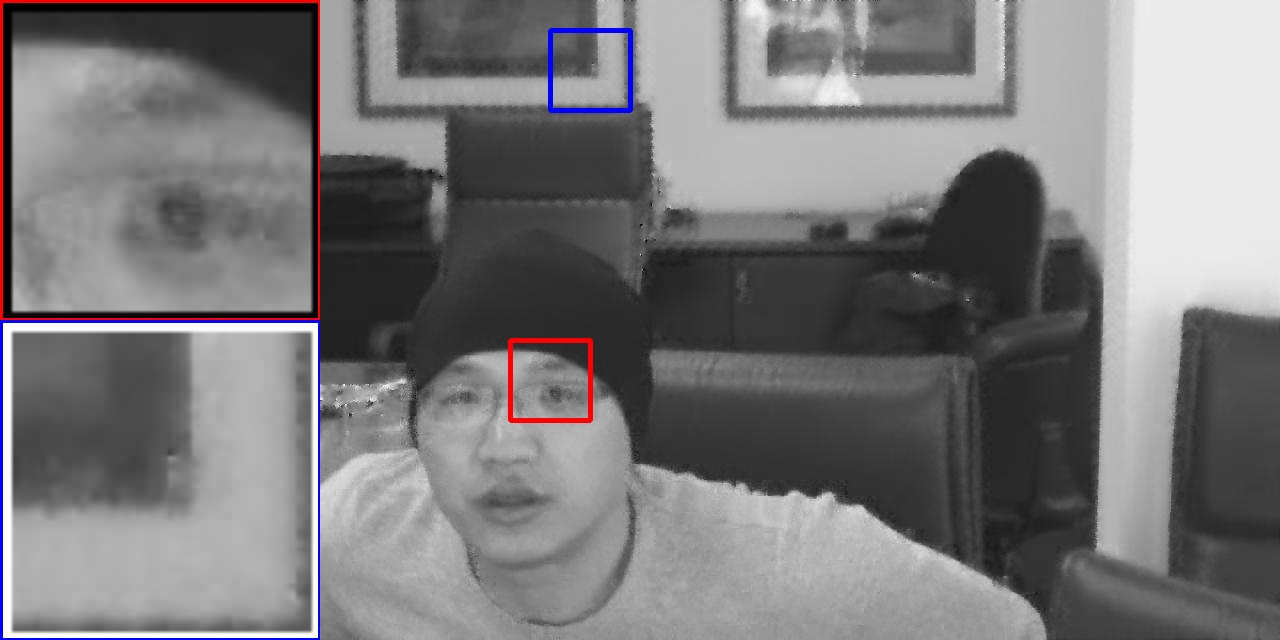}
      \caption*{MC-BCS-SPL\cite{5749476}\protect\\26.491/0.847}
       \includegraphics[width=\linewidth]{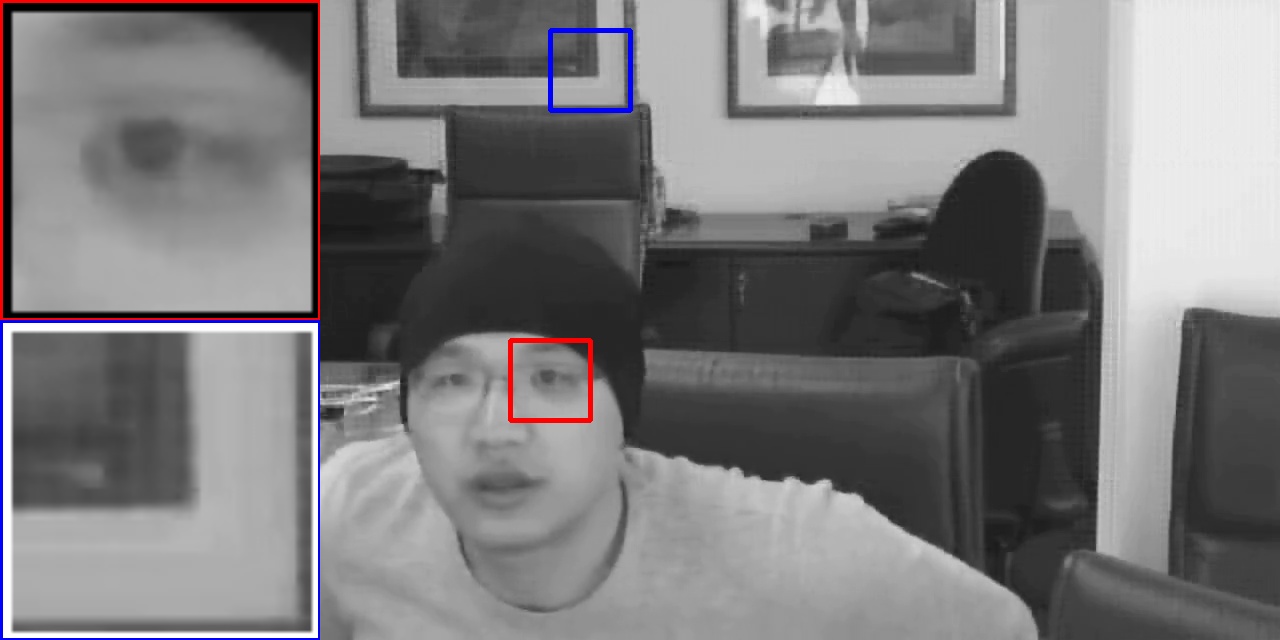}
      \caption*{CS-MCNet:{\it (proposed)}\protect\\27.379/0.880}
    \end{minipage}
  }
\end{minipage}
\vfill
\caption{Visual results and PSNR/SSIM metric of reconstructed frames of reference methods ISTANet\cite{8578294}, DeepVideoCS\cite{ILIADIS20189}, CSVideoNet\cite{8354291}, MC-BCS-SPL\cite{5749476}and the proposed approach. The original frames are also presented in the figure.}
\label{fig6}
\end{figure*}

Visual examples of some reference methods are shown in Fig. \ref{fig6}. The ground truth is also shown in Fig. \ref{fig6}. As we can see, CS-MCNet provides the best detail of selected methods and suffers minimal block effect. CS-MCNet produces sharper edges in the highlight areas and a more uniform image overall. This comparison demonstrates that inter-frame information is significant for video CS reconstruction, and the image CS approaches are not suitable for video tasks. Through further optimization, we believe that CS-MCNet has the potential to be applied on real-time reconstruction of high-frame-rate video CS.

\captionsetup[subfigure]{singlelinecheck=off,justification=raggedright}

\subsection{Reconstruction Under Noise}
In this subsection, we investigate the performance of our proposed network in the presence of measurement noise. The measurement of CS usually involves noise in practice caused by devices, and the measurement model should now be modified as

\begin{equation}
y = \Phi \cdot x + n\;.
\label{eq7}
\end{equation}

where n is the additive measurement noise.

\begin{figure}
\centering
\includegraphics[width=0.8\linewidth]{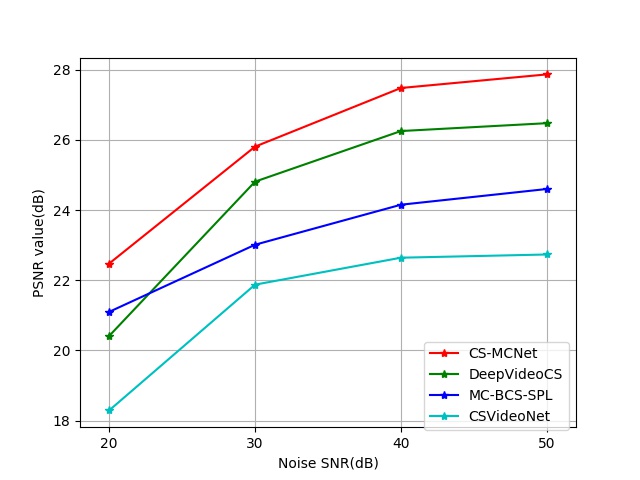}
\caption{Average PSNR over test dataset for several methods under different levels of measurement noise.}
\label{fig7}
\end{figure}

We conduct experiment with input measurements contaminated by random Gaussian noise, and all other parameters remain the same as in the noiseless case. We test the performance at four level of SNR from 20dB to 50dB under CR of 16. The result is shown in Fig. \ref{fig7}. It can be observed that at different noise level, CS-MCNet can achieve stable reconstruction performance and outperform the reference methods consistently. It is worth emphasizing that we did not retrain the network with noise measurement and all the experiments in this subsection are implemented with noiseless model, which shows the robustness of our model under different measurement conditions. Besides, it is easy to combat with performance degradation under noise by cascading our network with a deep denoising architecture or other denoising algorithm.

\subsection{Discussion}
As described earlier, the number of stages of CS-MCNet corresponds to the number of iterations of the original algorithm. In this subsection, we mainly focuses on the structure parameter of CS-MCNet, {\it i.e.} the nubmer of stages. From Table \ref{tab3}, we can find that as the number of stages increases from 2 to 4, the performance improves under different CRs. This can be explained by the fact that the deeper the neural network, the better its learning capacity. However, as the number of stages increases to 5, the performance deteriorates. We speculate that the reason is that while a deeper structure may help to fit the training data more accurately, it also makes it more difficult to train, resulting in an undertrained model. The training time increased slightly when the number of stages is less than 4 and increases rapidly when the number of stages is further increased. For CS-MCNet with no more than 4 stages, the time used to train an epoch varies from 30 to 40 minutes with GPU acceleration, but for networks with 5 or more stages, it can take more than an hour. To strike a balance between the effectiveness of the reconstruction and the cost of training, we empirically decides to use CS-MCNet with 4 stages.

\setlength{\tabcolsep}{4pt}
\begin{table}
\begin{center}
\caption{Performance comparison of proposed approach with different number of stages.}
\label{tab3}
\begin{tabular}{>{\raggedright\arraybackslash}p{40 pt} | >{\raggedright\arraybackslash}p{40 pt} | >{\raggedright\arraybackslash}p{50 pt} >{\raggedright\arraybackslash}p{50 pt} >{\raggedright\arraybackslash}p{50 pt} >{\raggedright\arraybackslash}p{40 pt}}
\noalign{\smallskip}
\hline
\noalign{\smallskip}
CR & Metric & 2 stages & 3 stages & 4 stages & 5 stages\\
\noalign{\smallskip}
\hline
\noalign{\smallskip}
4 & PSNR &32.776 &32.892& \it{32.936} & 32.656 \\
& SSIM & 0.904 & 0.905 & 0.909 & \it{0.910} \\
\noalign{\smallskip}
\hline
\noalign{\smallskip}
16 & PSNR &26.268 & 26.399 & \it{26.834} & 26.097\\
& SSIM &0.691 & 0.697 & \it{0.698} & 0.684\\
\noalign{\smallskip}
\hline
\noalign{\smallskip}
64 & PSNR &21.312 & 21.412 &\it{21.447} & 21.359 \\
& SSIM &0.379  &\it{0.391} &0.382 & 0.386 \\
\noalign{\smallskip}
\hline
\noalign{\smallskip}
Average & PSNR &26.864 & 26.862 &\it{26.979} &26.797 \\
& SSIM &0.658 &\it{0.664} &0.663& 0.659 \\
\hline
\end{tabular}
\end{center}
\end{table}
\setlength{\tabcolsep}{1.4pt}

\section{Conclusion}
Inspired by the MC-BCS-SPL algorithm, we use algorithmic unrolling to build a novel deep neural network to perform video compressive sensing reconstruction. Our proposed CS-MCNet has an interpretable multi-hypothesis motion compensation module that can exploit the correlation of neighboring frames in the video, which is important for improving reconstruction quality. The feedforward structure allows for fast CS reconstruction using GPU acceleration. CS-MCNet has been shown to outperform the reference method in terms of reconstruction quality and time consumption, and has the potential to be developed as a common framework for video CS applications. One direction of our future work is to integrate this network with video codec systems in general, and specifically on the task of bit rate control.

\bibliographystyle{splncs}
\bibliography{0771}

\begin{thebibliography}{10}

\bibitem{4016283}
{Candes}, E.J., {Tao}, T.:
\newblock Near-optimal signal recovery from random projections: Universal
  encoding strategies?
\newblock IEEE Transactions on Information Theory \textbf{52} (2006)
  5406--5425

\bibitem{1614066}
{Donoho}, D.L.:
\newblock Compressed sensing.
\newblock IEEE Transactions on Information Theory \textbf{52} (2006)
  1289--1306

\bibitem{4472247}
{Duarte}, M.F., {Davenport}, M.A., {Takhar}, D., {Laska}, J.N., {Sun}, T.,
  {Kelly}, K.F., {Baraniuk}, R.G.:
\newblock Single-pixel imaging via compressive sampling.
\newblock IEEE Signal Processing Magazine \textbf{25} (2008)  83--91

\bibitem{4472240}
{Candes}, E.J., {Wakin}, M.B.:
\newblock An introduction to compressive sampling.
\newblock IEEE Signal Processing Magazine \textbf{25} (2008)  21--30

\bibitem{doi:10.1002/cpa.20124}
Candès, E.J., Romberg, J.K., Tao, T.:
\newblock Stable signal recovery from incomplete and inaccurate measurements.
\newblock Communications on Pure and Applied Mathematics \textbf{59} (2006)
  1207--1223

\bibitem{Baraniuk2008ASP}
Baraniuk, R., Davenport, M.A., DeVore, R.A., Wakin, M.B.:
\newblock A simple proof of the restricted isometry property for random
  matrices.
\newblock Constructive Approximation \textbf{28} (2008)  253--263

\bibitem{8551508}
{Zhou}, J., {Zhou}, J., {Guo}, L.:
\newblock Angular intra prediction based measurement coding algorithm for
  compressively sensed image.
\newblock In: 2018 IEEE International Conference on Multimedia Expo Workshops
  (ICMEW). (2018)  1--6

\bibitem{6738005}
{Yang}, J., {Yuan}, X., {Liao}, X., {Llull}, P., {Sapiro}, G., {Brady}, D.J.,
  {Carin}, L.:
\newblock Gaussian mixture model for video compressive sensing.
\newblock In: 2013 IEEE International Conference on Image Processing. (2013)
  19--23

\bibitem{doi:10.1002/cpa.20042}
Daubechies, I., Defrise, M., De~Mol, C.:
\newblock An iterative thresholding algorithm for linear inverse problems with
  a sparsity constraint.
\newblock Communications on Pure and Applied Mathematics \textbf{57} (2004)
  1413--1457

\bibitem{4907073}
{He}, L., {Carin}, L.:
\newblock Exploiting structure in wavelet-based bayesian compressive sensing.
\newblock IEEE Transactions on Signal Processing \textbf{57} (2009)  3488--3497

\bibitem{7457256}
{Metzler}, C.A., {Maleki}, A., {Baraniuk}, R.G.:
\newblock From denoising to compressed sensing.
\newblock IEEE Transactions on Information Theory \textbf{62} (2016)
  5117--5144

\bibitem{QU2012964}
Qu, X., Guo, D., Ning, B., Hou, Y., Lin, Y., Cai, S., Chen, Z.:
\newblock Undersampled mri reconstruction with patch-based directional
  wavelets.
\newblock Magnetic Resonance Imaging \textbf{30} (2012)  964 -- 977

\bibitem{7337391}
{Zhan}, Z., {Cai}, J., {Guo}, D., {Liu}, Y., {Chen}, Z., {Qu}, X.:
\newblock Fast multiclass dictionaries learning with geometrical directions in
  mri reconstruction.
\newblock IEEE Transactions on Biomedical Engineering \textbf{63} (2016)
  1850--1861

\bibitem{5749476}
{Mun}, S., {Fowler}, J.E.:
\newblock Residual reconstruction for block-based compressed sensing of video.
\newblock In: 2011 Data Compression Conference. (2011)  183--192

\bibitem{8354291}
{Xu}, K., {Ren}, F.:
\newblock Csvideonet: A real-time end-to-end learning framework for
  high-frame-rate video compressive sensing.
\newblock In: 2018 IEEE Winter Conference on Applications of Computer Vision
  (WACV). (2018)  1680--1688

\bibitem{YAO2019483}
Yao, H., Dai, F., Zhang, S., Zhang, Y., Tian, Q., Xu, C.:
\newblock Dr2-net: Deep residual reconstruction network for image compressive
  sensing.
\newblock Neurocomputing \textbf{359} (2019)  483 -- 493

\bibitem{ILIADIS20189}
Iliadis, M., Spinoulas, L., Katsaggelos, A.K.:
\newblock Deep fully-connected networks for video compressive sensing.
\newblock Digital Signal Processing \textbf{72} (2018)  9 -- 18

\bibitem{7780424}
{Kulkarni}, K., {Lohit}, S., {Turaga}, P., {Kerviche}, R., {Ashok}, A.:
\newblock Reconnet: Non-iterative reconstruction of images from compressively
  sensed measurements.
\newblock In: 2016 IEEE Conference on Computer Vision and Pattern Recognition
  (CVPR). (2016)  449--458

\bibitem{8578294}
{Zhang}, J., {Ghanem}, B.:
\newblock Ista-net: Interpretable optimization-inspired deep network for image
  compressive sensing.
\newblock In: 2018 IEEE/CVF Conference on Computer Vision and Pattern
  Recognition. (2018)  1828--1837

\bibitem{10.5555/3104322.3104374}
Gregor, K., LeCun, Y.:
\newblock Learning fast approximations of sparse coding.
\newblock In: Proceedings of the 27th International Conference on International
  Conference on Machine Learning. ICML’10, Madison, WI, USA, Omnipress (2010)
   399–406

\bibitem{8550778}
{Yang}, Y., {Sun}, J., {Li}, H., {Xu}, Z.:
\newblock Admm-csnet: A deep learning approach for image compressive sensing.
\newblock IEEE Transactions on Pattern Analysis and Machine Intelligence
  \textbf{42} (2020)  521--538

\bibitem{erhan2010why}
Erhan, D., Bengio, Y., Courville, A., Manzagol, P., Vincent, P., Bengio, S.:
\newblock Why does unsupervised pre-training help deep learning?
\newblock Journal of Machine Learning Research \textbf{11} (2010)  625--660

\bibitem{10.5555/2843012}
Tekalp, A.M.:
\newblock Digital Video Processing. 2nd edn.
\newblock Prentice Hall Press, USA (2015)

\bibitem{1094950}
{Jain}, J., {Jain}, A.:
\newblock Displacement measurement and its application in interframe image
  coding.
\newblock IEEE Transactions on Communications \textbf{29} (1981)  1799--1808

\bibitem{5749477}
{Tramel}, E.W., {Fowler}, J.E.:
\newblock Video compressed sensing with multihypothesis.
\newblock In: 2011 Data Compression Conference. (2011)  193--202

\bibitem{7780459}
{He}, K., {Zhang}, X., {Ren}, S., {Sun}, J.:
\newblock Deep residual learning for image recognition.
\newblock In: 2016 IEEE Conference on Computer Vision and Pattern Recognition
  (CVPR). (2016)  770--778

\bibitem{Soomro2012UCF101AD}
Soomro, K., Zamir, A.R., Shah, M.:
\newblock Ucf101: A dataset of 101 human actions classes from videos in the
  wild.
\newblock ArXiv \textbf{abs/1212.0402} (2012)

\end{thebibliography}

\end{document}